# Method to Assess the Temporal Persistence of Potential Biometric Features: Application to Oculomotor, and Gait-Related Databases (August 2016)

Lee Friedman, Ioannis Rigas, *Member, IEEE*, Mark S. Nixon, and Oleg V. Komogortsev, *Member, IEEE*

***Abstract*— Although temporal persistence, or permanence, is a well understood requirement for optimal biometric features, there is no general agreement on how to assess temporal persistence. We suggest that the best way to assess temporal persistence is to perform a test-retest study, and assess test-retest reliability. For ratio-scale features that are normally distributed, this is best done using the Intraclass Correlation Coefficient (ICC). For 10 distinct data sets (8 eye-movement related, and 2 gait related), we calculated the test-retest reliability ("Temporal persistence") of each feature, and compared biometric performance of high-ICC features to lower ICC features, and to the set of all features. We demonstrate that using a subset of only high-ICC features produced superior Rank-1-Identification Rate (Rank-1-IR) performance in 9 of 10 databases (p = 0.01, one-tailed). For Equal Error Rate (EER), using a subset of only high-ICC features produced superior performance in 8 of 10 databases (p = 0.055, one-tailed). In general, then, prescreening potential biometric features, and choosing only highly reliable features will yield better performance than lower ICC features or than the set of all features combined. We hypothesize that this would likely be the case for any biometric modality where the features can be expressed as quantitative values on an interval or ratio scale, assuming an adequate number of relatively independent features.**

## I. Introduction

It has been stated that biometric features need to be permanent [1, 2]. However, "permanence" in the sense of never changing, does not actually apply: of course, no human feature is truly permanent. When the term permanence is used, frequently what is referred to is temporal persistence in the order of years, decades or a lifetime. We suggest that a more general term would be temporal persistence. All features need to be temporally persistent over the relevant Gallery-Probe test-retest time-frame. If this time-frame is minutes to months, the term permanence does not fit.

The assessment of temporal persistence is an element of all biometric types, but is perhaps much less of a problem when dealing with anatomically based biometric modalities (e.g., fingerprints, iris scans) than when dealing with physiological (e.g., electrocardiogram or electroencephalogram) or behavioral modalities (e.g., speech recognition, gait, or eye-movements).

The method for the assessment of temporal persistence is less obvious. Our view is that temporal persistence can only be assessed after one has conducted a test-retest reliability study, and that the appropriate quantitative metric indexing temporal persistence is a test-retest reliability estimate. For interval-scale or ratio-scale features that are normally distributed, the preferred statistic is the intraclass correlation coefficient, or ICC [3, 4]. For non-normally distributed features, or for ordinal or nominal scale features, other statistics, such as Cohen's Weighted Kappa, might be appropriate [5-9].

The ICC has several forms [3, 4], but for the present case, with two test intervals, and random subjects, the ICC for absolute agreement can simply be thought of as the variance due to subjects divided by the total variance. In this experiment, there are 3 sources of variance that can be distinguished: (1) variance due to subjects, (2) variance due to occasion, and (3) residual variance. Total variance is the sum of these 3 variances. Thus, the ICC can be thought of as the proportion of total variance that is due to subjects. It is highest when there is substantial subject variance, and minimal occasion, and residual variance. It ranges from 0.0 to 1.0. Cichetti, and Sparrow [10, 11] have suggested that ICC levels be interpreted by the following rules of thumb: ICC >= 0.75: "Excellent", ICC >= 0.60, and < 0.75: "Good", ICC >= 0.40, and < 0.60: "Fair", and

Submitted on July 30, 2016. This work was supported in part by NSF CAREER grant #CNS-1250718, and NIST grant #60NANB15D325.

L. Friedman is with the Department of Computer Science, Texas State University, 601 University Drive, San Marcos, Texas, 78666, USA (e-mail: lfriedman10@gmail.com).

I. Rigas is with the Department of Computer Science, Texas State University, 601 University Drive, San Marcos, Texas, 78666, USA (e-mail: rigas@txstate.edu).

M. S. Nixon is with the Department of Electronics, and Computer Science, University of Southampton, United Kingdom. (email: msn@ecs.soton.ac.uk).

O.V. Komogortsev is with the Department of Computer Science, Texas State University, 601 University Drive, San Marcos, Texas, 78666, USA (e-mail: ok@txstate.edu).

ICC < 0.40: "Poor".

This paper introduces the notion of temporal persistence in biometrics, enumerates it using the intraclass correlation coefficient, and evaluates it on numerous databases to demonstrate validity, and efficacy of this new notion.

## II. Methods

### A. Databases--Overview

We report on 10 databases. All data sets are briefly outlined here, and explained in detail below. There are 8 eye-movement related databases (EM-1-Short-Term, EM-1-Long-Term, EM-2-Short-Term, EM-2-Long-Term, EM-3-Short-Term, EM-3-Long-Term, EM-4-Short-Term, and EM-4-Long-Term), and 2 gait-related databases (Gait-1, and Gait-2). Four key databases are presented in the main manuscript (EM-1-Short-Term, EM-1-Long-Term, Gait-1, and Gait-2), and 6 additional databases are presented in the Supplementary Material (Supplementary_Material: EM-2-Short-Term, EM-2-Long-Term, EM-3-Short-Term, EM-3-Long-Term, EM-4-Short-Term, EM-4-Long-Term,). All the short-term EM databases are based on the same subjects, using the same eye-movement recordings, during the same task. In this case, the test-retest interval is on the order of 19 minutes (within this report, repeated testing periods on the same subjects within a day are referred to as "sessions"). All the long-term EM databases are based on the same subjects, using the same eye-movement recordings, during the same task. In this case, the test-retest interval is on the order of 11.1 months (range: 7.8 to 13.0 months). The eye-movement data were collected as part of a larger database collected at Texas State University as a result of the NSF CAREER grant awarded to Dr. Komogortsev. The EM databases differ, however, in the sense that different, but not necessarily mutually independent, sets of features were extracted from eye-movement signals. The subjects were reading a poem for up to 60 seconds. EM-1 is a new database, based on a new comprehensive feature set never described before in print, but described in some detail below. EM-2 is based on so called "Complex Eye Movement" feature set described in [12]. EM-3 is based on oculomotor plant characteristics (OPC) for biometric identification, as described in [13]. EM-4 is based on so called "Complex Eye Movement Pattern" biometrics, as described in [14]. Both gait-related databases employ the Southampton Large Population Gait database of gait-related images, and videos [15]. These databases are comprised of over 100 subjects tested on many sessions. Generally, the gait assessments analyzed herein were all collected during the same day. In the present analysis, we report only on the Session 1 vs Session 2 comparison. Gait-1 employs model-based features extracted from these images [16]. Gait-2 is based on Zernike velocity moments extracted from these images [17]. Note that these gait-related databases, and analyses were chosen to illustrate the importance of the assessment of temporal persistence. They do not represent the best performance achievable with gait-related databases.

### B. Initial Feature Set Reduction

Our first step for all databases was to determine which features were normally distributed or could be transformed to normality using variable transformations. This was done using all data from all subjects from the first session. We wanted to assess the reliability of each measure using the ICC, which requires that the data be normally distributed. We assessed normality using the Pearson Chi-Square test. For measures that were not normally distributed, standard transformations were applied to the data (reciprocal, log, square root, cube root, reflected, logit, arcsine, winsorization [18, 19]). Measures which were not normally distributed, and could not be transformed into normal, were dropped from further analysis.

### C. Removal of Redundant Features

For some databases, there were obviously redundant features. For example, some features were based on either the mean or the median of the same distribution, or the standard deviation (SD) or the interquartile range (IQR) of the same distribution. We did not need two estimates of the central tendency of eye-movement feature distributions (mean, and median) so the less reliable (lowest ICC) measure was dropped from further analysis. Similarly, we did not need two measures of variance [interquartile range (IQR), and standard deviation (SD)] so the less reliable was dropped. We also intercorrelated every feature with every other feature, and found those pairs of features that were intercorrelated (Pearson's r) greater than 0.95 (absolute value). We considered such pairs of features effectively redundant. The lower reliability feature from each pair was dropped from further analysis. This removal of redundant features was done for all subjects for session 1 data.

### D. Calculation of the Intraclass Correlation Coefficient (ICC)

For each database, for each feature, we have data from N subjects for 2 sessions. (ICCs were calculated using all data from all subjects). From these data, we can estimate 3 variance components: Variance due to subjects, variance due to occasion (e.g., Session 1 vs Session 2), and error or residual variance. These variance components can be estimated indirectly from an ANOVA table [3, 4] or more directly, using a variance component analysis (lme4, R package [20]) (See Table I for details regarding the computation of the ICC). Since there are a number of different ICCs for different data structures, we recommend that, prior to implementing an ICC, readers consult references [3, 4], which describe the types of ICC and the basis for making a choice among them.

### E. Splitting the Features into ICC Sets

When there were many features (> 200, EM-1) we split the features into 3 equally sized ICC sets (High, Moderate, Low). When there were fewer features (EM-2, EM-3, EM-4, Gait-1, and Gait-2) the features were split into 2 equally sized ICC sets (High, Low). In addition, for all datasets, an "ALL" ICC set, containing all features, regardless of ICC, was also considered.

| Table I - How to Calculate the ICC | |
|---|---|
| **Requires Estimates of Variance Components** | |
| VarS | Variance due to Subjects |
| VarO | Variance due to Occasion |
| VarR | Residual Variance |
| ICC = | VarS/(VarS+VarO+VarR) |

| How to Estimate Variance Components | |
|---|---|
| (1) Variance Components Analysis | |
| (2) ANOVA Table (2-way, Random Effects): | |
| BMS | Mean Square Between Subjects |
| RMS | Mean Square Residual (Interaction + Error) |
| OMS | Mean Square Occasion |
| k | Number of Occasions |
| n | Number of Subjects |
| Numerator | BMS-RMS |
| Denominator | (BMS+(k-1)*RMS+k*(OMS-RMS))/n |
| ICC | Numerator/Denominator |

### *F. Assessing the Positive-Definiteness of ICC Data Sets*

During the assessment of biometric performance (see below), the data were separated into training, and testing sets, each containing N/2 subjects. PCA analysis was performed on each training set. PCA requires correlation matrices as input that are positive-definite (all eigenvalues > 0.0). Sometimes, the PCA algorithm would fail due to the non-positive definiteness of the relevant correlation matrix. In this case, we searched for the most highly intercorrelated pairs of features, and removed the member of a pair with a lower ICC value. We removed as few features as necessary to get the PCA to work without error.

### *G. Assessing Biometric Performance*

In every case the values of the features were directly used to form the corresponding feature vectors that were employed as the biometric templates. The comparison (matching) of biometric templates was performed via the use of the Cosine distance, with the resulting distances converted to similarity scores, and used to assess biometric performance. It should be noted that several other types of distances were tested (Euclidean, City-block, Spearman, Mahalanobis), and provided competitive performance. Since analyses based on the Cosine distance performed marginally better than the other distance metrics, the results we present here are based on that metric.

Biometric performance was assessed in terms of Rank-1 Identification Rate (Rank-1-IR), and equal error rate (EER).

Equal Error Rate (EER): The EER is a measure of the verification accuracy of a biometric system. A genuine score is defined as the score from the comparison of the biometric samples coming from the same identity. An impostor score is defined as the score from the comparison of biometric samples coming from different identities. By defining an acceptance threshold ($\eta$) we can compute the False Rejection Rate (FRR) as the percentage of the genuine scores that fall under the threshold $\eta$, and the False Acceptance Rate (FAR) as the percentage of the impostor scores that are over $\eta$. True Positive Rate (TPR) can be defined as the percentage of genuine scores that are over the threshold $\eta$, with TPR = 1 - FRR. By changing the acceptance threshold, we can construct a Receiver Operating Characteristic (ROC) curve, and calculate the EER as the point of operation where the FRR equals the FAR.

Rank-1 Identification Rate (Rank-1-IR): The Rank-k Identification Rate (Rank-k-IR) is a measure of biometric identification performance which shows the percentage of genuine scores that can be found within the k top places of a ranked list. A Cumulative Match Characteristic (CMC) curve shows the change of the identification rate as a function of the used rank k. The Rank-1-IR is defined as the percent of biometric samples with a correct match in the first place of the ranked list.

Biometric performance was assessed at the level of an ICC set [HIGH, MOD (EM-1), LOW, ALL]. Each ICC set consisted of k features for N subjects for 2 sessions. For each ICC set, the following procedures were followed: The set of N subjects was split into a training set (N/2 subjects), and a test set (N/2 subjects) using random sampling, without replacement. A principal component analysis (PCA) was conducted for dimension reduction, extracting from 2 to p components, where p is determined by the data set, and ranged from 6 for the EM-3-Short-Term data to 21 for the EM-1-Short-Term data. Components were extracted until a peak Rank-1-IR performance was achieved. Using the component structure from the training set, we created component scores for the test sample, both Session 1, and Session 2. These testing data sets were submitted to a biometric performance analysis treating Session 1 data as the gallery, and Session 2 data as the probe. We present the median Rank-1-IR, and the EER over the 100 random test samples for from 2 to p components. For EM-4 only, since there were only 10 total features, 5 in the HIGH ICC group, and 5 in the LOW ICC group, no PCA was conducted. All 5 or 10 features were entered into the biometric assessment algorithm directly.

### *H. EM-1 Database*

This analysis, and database have not been described in print before, so a more extensive discussion of the database will be presented for this database only.

The analysis presented here is the result of many important prior steps, including subject recruitment, eye-movement recording, eye-movement classification, feature extraction, assessment of normality of feature distributions, reliability assessment, feature redundancy elimination, and biometric assessment. These phases are being described in detail in separate publications in preparation. Therefore, in this manuscript, the prior steps will be described only briefly. In every case the values of the features were directly used to form the corresponding feature vectors that were employed as the biometric templates. The comparison (matching) of biometric templates was performed via the use of the Cosine distance, with the resulting distances converted to similarity

### I. EM-1: Subjects

The subjects were all undergraduate college students at Texas State University (N=333, N Female = 157, N Male = 178, mean age = 21.8, range = 18 to 46 yrs., EM-1-Short-Term). All materials, and procedures were approved by the Institutional Review Board at Texas State University, and informed consent was obtained from all participants before testing.

### J. EM-1: Eye-Movement Stimulus

On each subject visit, subjects were studied twice (Sessions 1, and 2), approximately 20 min apart (EM-1-Short-Term). Each Session included 7 different sets of visual stimuli, only 1 of which (Poetry Reading) is relevant to the present report. For each instance of the Poetry Reading task, a different pair of quatrains from the famous nonsense poem, “Hunting for a Snark”, written by Lewis Carroll (written from 1874-1876), were displayed. Subjects were asked to read the poem portion silently. They were given 60 seconds to read this poem passage. Session 1 to Session 2 (task-to-task) time intervals ranged from 13 min to 42 min (mean: 19.5; SD: 4.2, (EM-1-Short-Term)).

### K. EM-1: Eye-Movement Recording

The subjects were seated 55 cm in front of a computer monitor with their heads resting on a chin/head rest. The monitor subtended +/- 23.3 degrees of visual angle in the horizontal direction, 11.7 degrees to the top, and 18.5 degrees to the bottom. The Eyelink 1000 (SR Research Ltd., Ottawa, ON, Canada) was employed for eye-movement recording. It is video-oculography system, which records both horizontal, and vertical eye movements. The sampling rate for our data was 1000 Hz. In the present study, only left eye movements were collected. The device has an accuracy of 0.25 - 0.5 deg. For further specifications, see [22].

### L. EM-1: Eye-Movement Classification

In the present study, all eye movements were analyzed off-line. We began our classification of the eye movements using the method described by Nystrőm, and Holmqvist [23], using Matlab code kindly made available by Dr. Nystrőm. Over the course of many months, we modified this code so extensively to enhance its performance for our data that what we have ended up with is an effectively new eye-movement classification scheme, based on the general approach outlined in [23]. We are preparing a manuscript for publication describing the many changes we made to this algorithm, and assessing the performance of the new algorithm in comparison to the original code.

Briefly, the algorithm classifies eye-movement signals into fixation periods, saccades, glissades, and noise. It relies most heavily on the velocity trace, and largely ignores position, and acceleration for classification purposes. We only analyzed that portion of the signal where, according to our analysis, subjects were actually reading the poem. Most of the subjects took the entire 60 sec. to read the poem, but many subjects finished early.

Ten subjects did not have both Session 1, and Session 2 recordings, and twenty-five subjects were dropped from the study due to low recording quality. This left 298 subjects (Table II).

### M. EM-1: Eye-Movement Feature Extraction

Our overall philosophy with regard to feature extraction was to extract every conceivable, quantifiable, and objective feature we could imagine. The idea was that this very large set of features would be winnowed down to a reasonable number of features in several steps. We did not try to guess a priori which features would be useful, and which would not be useful, but rather determined the usefulness of the features empirically. Many features were obviously redundant (e.g., means, and medians), but this redundancy was removed in a later stage. For most measures, features were separately extracted for the horizontal component, the vertical component, and a radial component.

Most of the features were based on distributions of measures for each event type (fixation, saccade, and glissade), and for each such measure we extracted the mean, median, standard deviation, interquartile range, skewness, and kurtosis.

For fixations, we measured duration, rate (per second), fixation drift, velocity noise, shape in the position channel, shape in the velocity channel, and shape in the acceleration channel. For saccades, we measured duration, rate, shape in the positon channel, shape in the velocity channel, shape in the acceleration channel, peak velocity, peak acceleration, saccade trajectory curvature in the 2-dimensional plane (horizontal, and vertical), as well as main sequence relationships. For glissades, we measured size, percentage of saccades with glissades, shape in the positon channel, and shape in the velocity channel, shape in the acceleration channel, peak velocity, and peak acceleration. We also evaluated reading speed, number of small saccades (< 8 deg) to the right (presumably word-to-word saccades), number of small saccades to the left (presumably refixation saccades), and number of line-returning saccades. Finally, we also measured types, and character of artifacts, as well as pupil size. A technical report providing a detailed presentation of the construction of the various features is available upon request. As a result of feature extraction, we had 1027 potential eye-movement features for each subject, for each Session. After removal of features which were not normal, and could not be transformed into normal, and after removal of redundant features, 321 features remained (EM-1-Short-Term, Table II). These were further subdivided, based on ICC, into 3 ICC sets (HIGH, MOD, LOW), and an ALL dataset containing all 321 features. See Table II for sample sizes, test-retest interval, number of features, number of ICC sets, ICC set sizes, median ICC, and relevant cut-points for ICC sets.

**Table II. Description of Databases**

| Analysis | Reference | Number of Subjects | Test-Retest Interval | Number of Original Features | Number of Normal or Normalized Non-Redundant Features* | Number of ICC Sets | Median ICC | Cut Points |
|---|---|---|---|---|---|---|---|---|
| EM-1-ST | N/A | 298 | Median 19 min | 1047 | 321 (107) | 3 | 0.55 | 0.40, 0.72 |
| EM-1-LT | N/A | 68 | Median 11.1 mon | 1047 | 501 (167) | 3 | 0.41 | 0.23, 0.54 |
| EM-2-ST | 12 | 298 | Median 19 min | 84 | 54 (27) | 2 | 0.56 | NA |
| EM-2-LT | 12 | 68 | Median 11.1 mon | 84 | 48 (24) | 2 | 0.46 | NA |
| EM-3-ST | 13 | 298 | Median 19 min | 18 | 18 (9) | 2 | 0.49 | NA |
| EM-3-LT | 13 | 68 | Median 11.1 mon | 18 | 18 (9) | 2 | 0.43 | NA |
| EM-4-ST | 14 | 298 | Median 19 min | 14 | 10 (5) | 2 | 0.79 | NA |
| EM-4-LT | 14 | 68 | Median 11.1 mon | 14 | 12 (6) | 2 | 0.46 | NA |
| Gait-1 | 15, 16, 17 | 112 | Within Same Day | 74 | 70 (35) | 2 | 0.72 | NA |
| Gait-2 | 15, 16, 18 | 107 | Within Same Day | 31 | 30 (15) | 2 | 0.87 | NA |

* The number per ICC Set is in parentheses.

### *N. EM-1-Long-Term*

The data are of exactly the same form as the EM-1-Short-Term database, but the subjects are a subset of 78 subjects retested at approximately 11.1 months. Ten of the subjects had excessively noisy recordings, and so 68 subjects were analyzed. See Table II for sample sizes, test-retest interval, number of features, number of ICC sets, ICC set sizes, median ICC, and relevant cut-points for ICC.

For a description of EM-2, EM-3, and EM-4 Databases, see: Supplementary_Material.

### *O. Gait-1*

This database was was from the Southampton Large Population Gait database [15]. It is based on model-based features extracted from this database [16]. From the overall database, we selected 112 subjects with complete data for exactly 2 sessions. The database contained 74 model-based metrics (features), 70 of which were normal or could be normalized by transformation. See Table II for sample sizes, test-retest interval, number of features, number of ICC sets, ICC set sizes, and median ICC.

### *P. Gait-2*

This database was also from the Southampton Large Population Gait database [15]. This database is based on Zernike velocity moments [17]. We chose 107 subjects with complete data for 2 sessions. The Gait-2 database includes 31 features, 30 of which were normal or could be normalized. All ICCs for this database were in the excellent range (Table II). This is an extraordinarily reliable database. The 15 features with the highest ICC, based on a median-split, were combined in the High ICC set, and the 15 features with the lowest ICC were combined in the LOW ICC set. A third database, containing all 30 features was also created (“ALL”). See Table II for sample sizes, test-retest interval, number of features, number of ICC sets, ICC set sizes, and median ICC.

## III. Results

### *A. EM-1*

Short-Term: The analyses for this database are presented in Fig. 1, and Table III. There was a large range of ICCs for these 321 features (Fig. 1, Top). For all sets, performance generally improved as the number of components extracted increased to

**Table III. Biometric Performance Summary**

| Analysis | Rank 1 - IR | | | EER | | |
|---|---|---|---|---|---|---|
| | Best Set | Highest Rank 1 - IR | Number of Components | Best Set | Lowest EER | Number of Components |
| EM-1-ST | HIGH | 89.9 | 21 | HIGH | 2.68 | 19 |
| EM-1-LT | HIGH | 70.6 | 14 | HIGH | 12.00 | 10 |
| EM-2-ST | HIGH | 48.3 | 13 | HIGH | 11.41 | 10 |
| EM-2-LT | HIGH | 47.1 | 11 | HIGH | 16.00 | 7 |
| EM-3-ST | HIGH | 10.7 | 6 | HIGH | 21.45 | 3 |
| EM-3-LT | HIGH | 26.5 | 8 | ALL | 24.90 | 3 |
| EM-4-ST | ALL | 34.9 | N/A * | ALL | 15.40 | No PCA (see text) |
| EM-4-LT | HIGH | 50 | N/A * | HIGH | 17.00 | No PCA (see text) |
| Gait-1 | HIGH | 80.4 | 15 | HIGH | 5.36 | 11 |
| Gait-2 | Tie: HIGH & ALL | 81.1 | 9 | HIGH | 5.66 | 9 |

21. For the HIGH ICC set, Rank-1-IR performance was very good (Fig. 1, Middle, and Table III). For EER, the HIGH ICC set also performed very well (Fig. 1, Bottom, Table III). The performance of the HIGH ICC set was dramatically better than the performance of the other sets, including the ALL set, which contained more features by a factor of 3.

Long-Term: The analyses for this database are presented in Fig. 2, and Table III. There is a large range of ICCs for these 501 features (Fig. 2, Top). Note the decline in median ICC compared to EM-1-Short-Term (0.55 to 0.41). We expect ICC to decrease as a function of the length of the test-retest interval. For all sets, performance generally improved as the number of components extracted increased to 14. For the HIGH ICC set, Rank-1-IR performance was good (Fig. 2, Middle, and Table IV), although not as good as for the EM-1-Short-Term database (Rank 1-IR: 89.9 to 70.6). For EER, the HIGH ICC set also performed better than any other dataset, although the performance was not excellent (Fig. 2, Bottom, Table IV). The performance of the HIGH ICC set was dramatically better than the performance of the other sets, including the ALL set, which contained more features by a factor of 3.

For the results for EM-2, EM-3, and EM-4 databases, see: Supplementary_Material.

### B. *Gait-1 Database*

The analyses for this database are presented in Fig. 3, and Table III. There were 70 features for this dataset (Table II), and the ICCs were generally high (Fig. 3, Top). For all sets, for Rank-1-IR, performance generally improved as the number of components extracted increased to 15. The HIGH ICC set performed substantially better than the other ICC sets for both Rank-1-IR (Fig. 3, Middle), and EER (Fig. 3, Bottom). Overall, the performance was reasonably good for this database.

### C. *Gait-2 Database*

The analyses for this database are presented in Fig. 4, and Table IV. There were 30 features for this dataset (Table II), and the ICCs were all in the excellent range. This was a very reliable set of features. The HIGH ICC set, and the ALL ICC set both achieve identical Rank-1-IR peak performance at 9 components (Fig. 4, Middle). For EER, the best performance was for both the HIGH-ICC set, and the ALL set, but the HIGH ICC set does fractionally better (Fig. 4, Bottom, Table IV). Overall, the performance was reasonably good for this database.

### D. *Tests of Probability of Proportions*

In 8 of 10 cases, the Rank-1-IR for the HIGH-ICC set was larger than other datasets. In one case, the HIGH ICC set was tied for performance with the ALL ICC set. Given a tie, we can score this also as a win for the HIGH-ICC set, since it accomplishes the same performance with 50% of the measures. This amounts to a proportion of 9/10, or 0.90. The probability of getting a proportion this extreme or more extreme, when the null hypothesis is chance (probability = 0.50) is p = 0.011 (one-tailed, Exact Binomial Test). For EER, the HIGH-ICC was lower than other ICC sets in 8/10 cases (p = 0.055, one-tailed, Exact Binomial Test).

## IV. Discussion, and Conclusion

We have shown that the ICC carries information that is highly relevant to biometrics. We propose to refer to this information as temporal persistence. For Rank-1-IR, in 9 of 10 (p = 0.011) databases, the HIGH ICC set of features performed better than LOW ICC sets of features, and the MOD (moderate) ICC set of features. For EER, in 8 of 10 (p = 0.055) databases, the HIGH ICC set of features performed better than LOW ICC sets of features, and the MOD (moderate) ICC set of features. Furthermore, the MOD ICC set of features performed better than the LOW ICC set for the databases where it was created as a separate group (EM-1-Short-Term, EM-1-Long-Term).

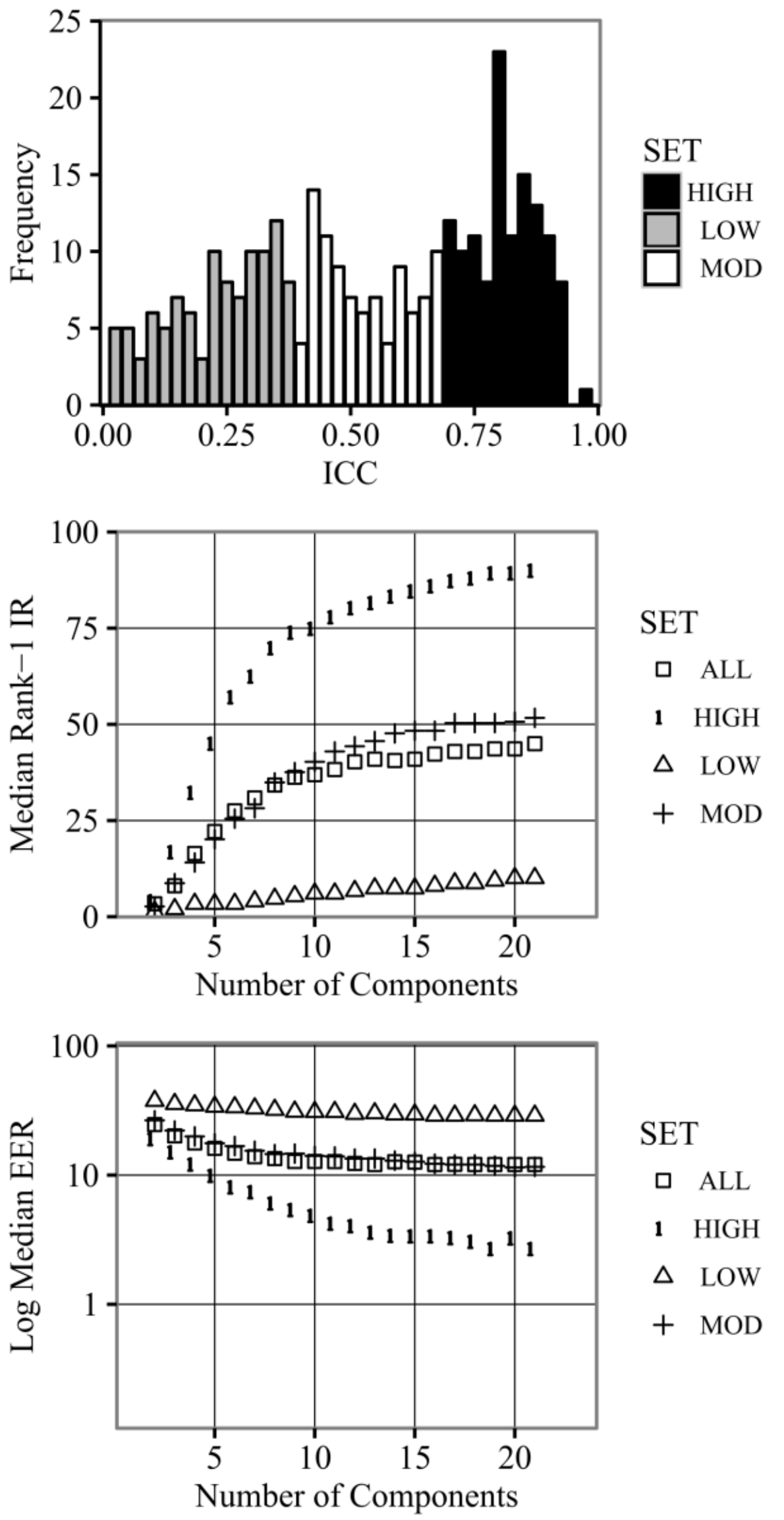

Figure 1. Analysis of the EM-1-Short-Term database. Top plot: ICC Histogram. Low ICC features (gray bars), moderate ICC features (white bars), high ICC features (black bars). Middle plot presents the Rank-1-IR results as a function of the number of PCA components. Lower plot presents the EER results. Each symbol represents the median over 100 random training-testing sets.

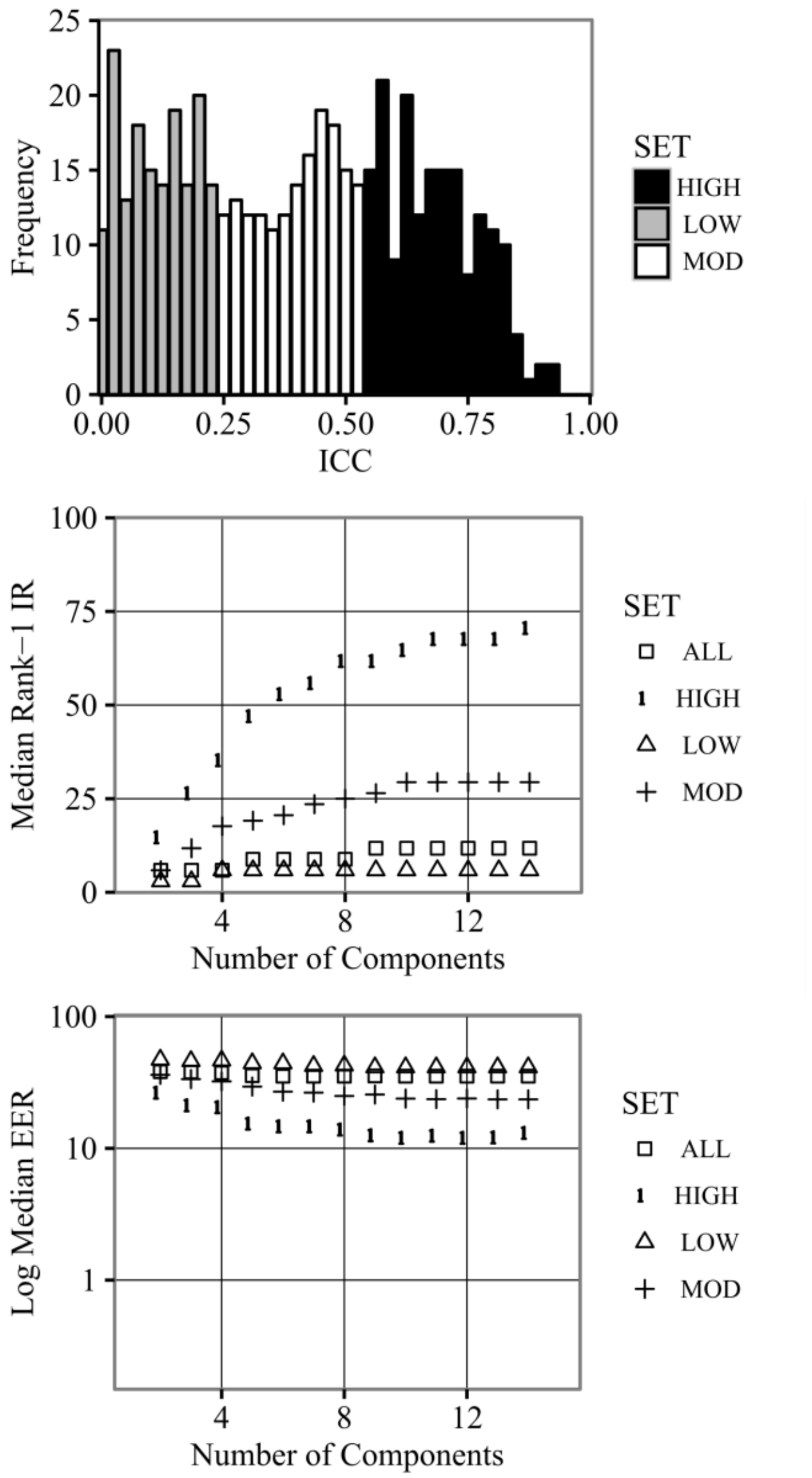

Figure 2. Analysis of the EM-1-Long-Term database. See caption for Fig. 1.

The only competition to the excellent performance of the HIGH ICC set was from the ALL ICC groups. What is being compared here is one set, the HIGH ICC set, as a subset of another set, the ALL ICC set. In 8 of 10 databases, this means that the HIGH ICC set employed 50% of the features in the ALL ICC set, and in 2 databases, the HIGH ICC set had only 33% of the features in the ALL ICC set. The implication of this finding is that, not only were HIGH ICC measures better, but that even including lower ICC features with high ICC features leads to decreased biometric performance. So this not only supports our view on the strength, and importance of the ICC as a measure of temporal persistence for biometric analysis, but actually supports our specific strategy of dividing the features into different ICC sets, and only using the highest ICC set of features.

In one case, the Rank-1-IR for the HIGH ICC set, and the ALL ICC set were tied (Gait-2). In this case, all the features had very high ICC (range, 0.75 to 1.0, median = 0.87). This was an extraordinarily reliable set of features. This finding suggests the obvious conclusion that the value of sorting features on the basis of relative temporal persistence is going to be a function of the range, and spread of temporal persistence estimates (ICCs) across the features.

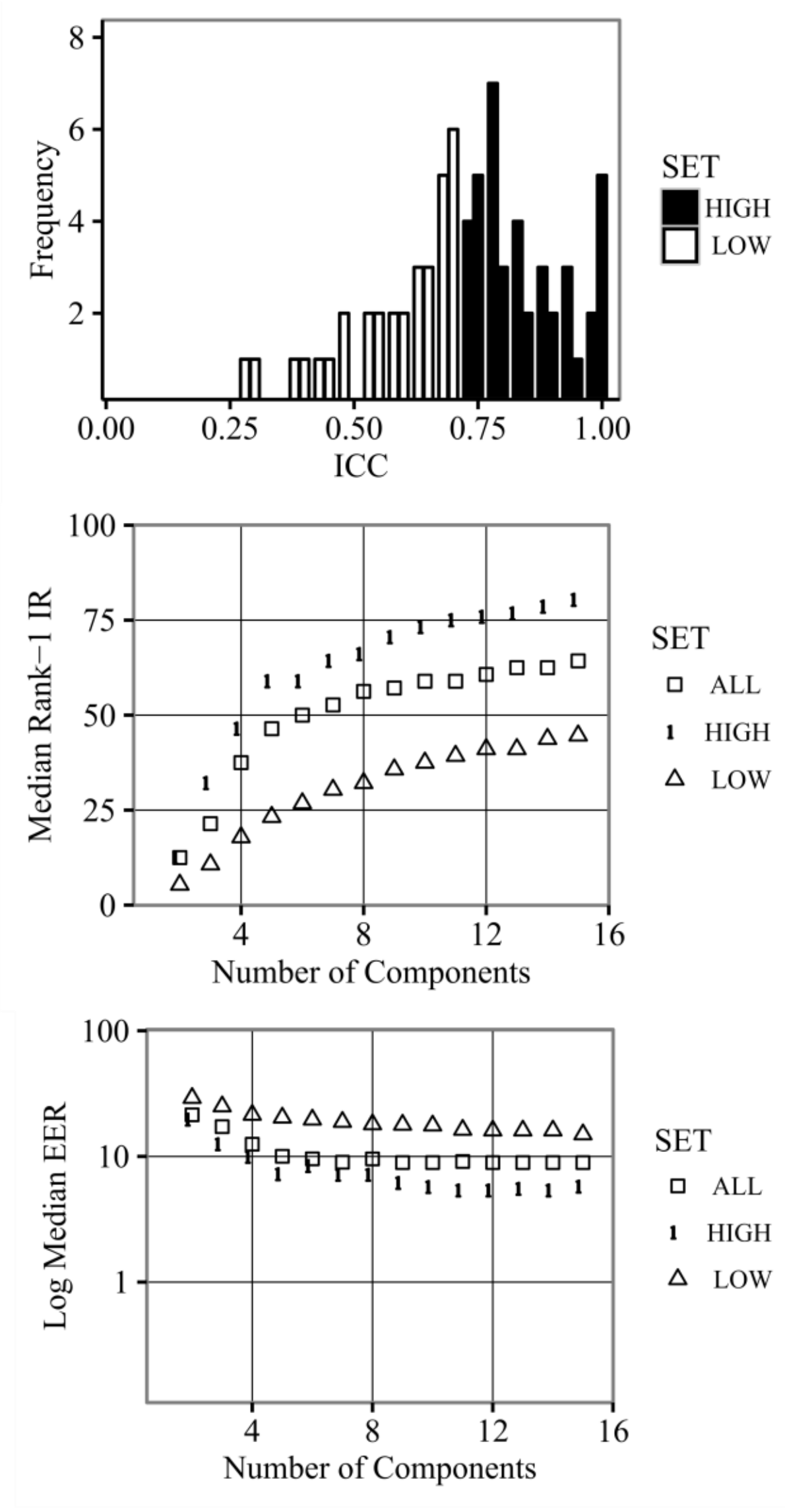

Figure 3. Analysis of the Gait-1 database. See caption for Fig. 1.

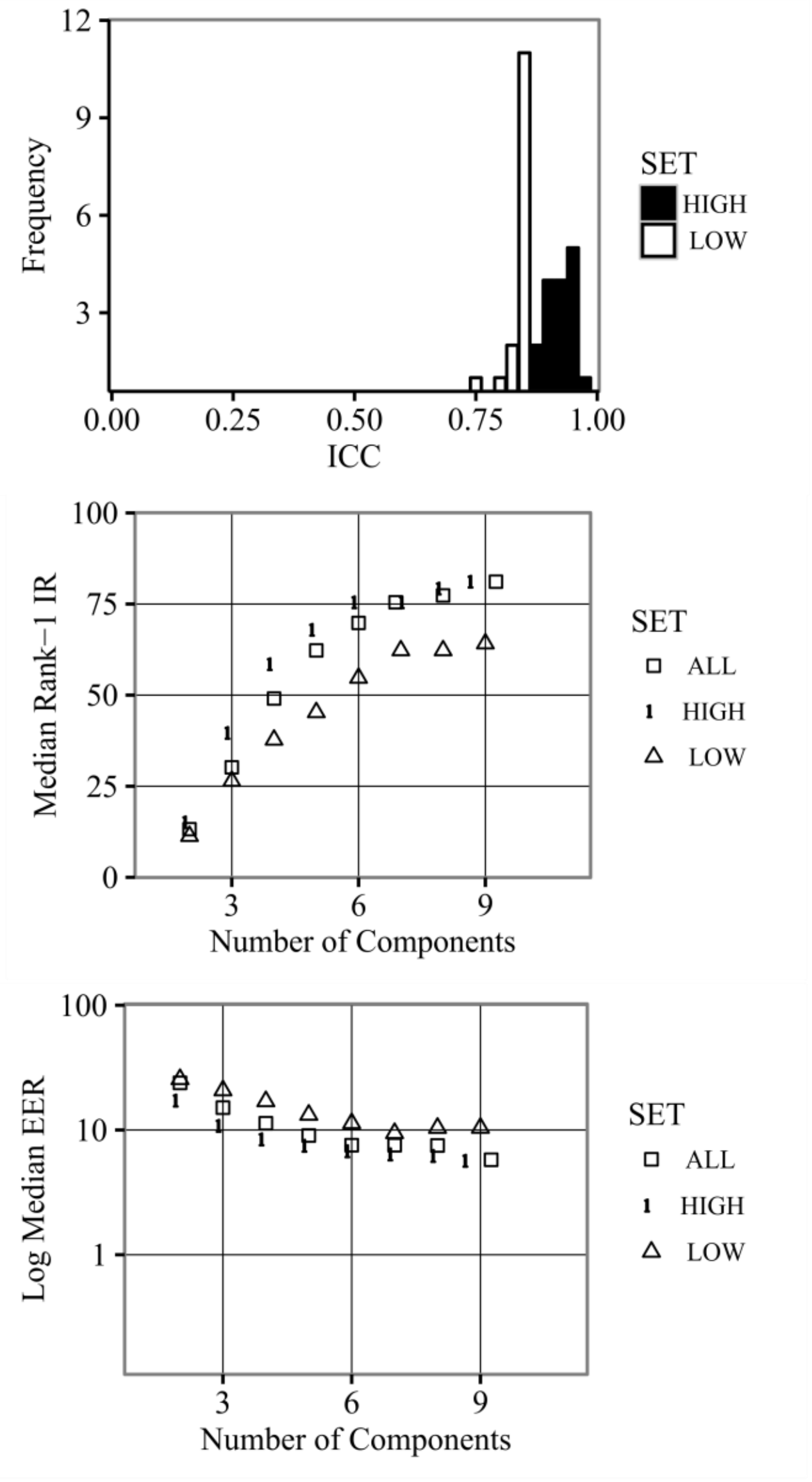

Figure 4. Analysis of the Gait-2 database. See caption for Fig. 1.

Table IV. Published Performance for other EM-Related Approaches from Our Laboratory

| Authors | Year | Reference | Approach | Best EER | Best Rank-1 IR |
|---|---|---|---|---|---|
| I. Rigas, O.V. Komogortsev | 2014 | [30] | Fixation Density Map | 10.8 | 51.0 % |
| O.V. Komogortsev, A. Karpov, and C. Holland | 2016 | [31] | Oculomotor Plant Characteristics | 14.0 | 24.7 % |
| I. Rigas, O.V. Komogortsev, and R. Shadmehr | 2016 | [32] | CEM-B + Acceleration + Vigor | 10.6 | 64.3 % |
| I. Rigas, E. Abdulin, and O.V. Komogortsev | 2016 | [21] | Multi-Source Fusion | 5.8 | 88.6 % |
| Present results for EM-1-Short-Term | | | HIGH ICC set (107 features, 21 PCA components) | 2.7 | 89.9% |

In one case (EM-4), the ALL set clearly outperformed the HIGH ICC set. In this case, the database only contained 10 normal features. This result reminds us that a high degree of temporal persistence is not the only important quality for potential biometric features. Obviously, one must have a sufficient number of independently informative features.

We would like to invite any reader who might have a database that might be tested with the approach presented herein, to either conduct this analysis themselves or send it to us for analysis. At a minimum, we will return the results to the contributor, and beyond that we can discuss making all such databases, and results available on a public website and/or co-publish the findings.

The biometric performance achieved with the HIGH ICC set from the EM-1-Short-Term database was the best we have achieved in our laboratory (Table IV). Although we came close to the present results with our Multi-Source Fusion technique [21], it is important to keep in mind that the analysis of Rigas *et al.*, [21] was based on much more data, was much more complicated, and involved fusion at the stimulus level (Poetry reading, and other tasks), and fusion of multiple biometric algorithms.

We are not aware of any prior biometric study which employed a test-retest paradigm, and evaluated the ICC as an index of temporal persistence of potential features.

It is somewhat surprising that prior biometric researchers have not employed the ICC to assess temporal persistence, since, in our view, this coefficient is high suitable for biometrics. Furthermore, it was introduced by R.A. Fisher in 1925 [22], and it has been employed in nearly 8,000 published studies according to the National Library of Medicine (PubMed).

We have found one paper that employed the Pearson's r to assess repeatability [23], and another paper that used the t-value for a regression of time 2 onto time 1 [24]. The Pearson's r is an *inter*class correlation coefficient, whereas the ICC is an *intra*class correlation coefficient. The former is invariant to linear transformation of the time 2 data with respect to the time 1 data, whereas the ICC is a measure of absolute, not relative, agreement. Anything which prevents one from getting exactly the same score will reduce the ICC but may not affect Pearson's r. The same can be said of a t-value from a linear regression, i.e., this is also invariant to linear transformations. Another study created their own "stability index" [25] where it appears that an ICC would have been more appropriate. For assessments on a nominal scale, prior researchers have employed percentage agreement between time 1, and time 2 [26]. As noted by the authors as well as Bartko [27], and others, this procedure ignores chance agreements, and is considered a less suitable measure than Kappa [5]. One study employed an interesting meta-analytic technique to assess the reliability of nominal assessments by fingerprint experts [28].

We found one paper [29] that employed the ICC in the context of feature selection for face recognition biometrics. These authors did not employ the ICC to assess temporal persistence. The data structure in that paper differed from the data structure in the present case. In that paper, there were multiple different face images for each subject. The ICC was used to find features which were consistent across all the faces for each individual.

The median reliability of the gait-related features was much higher than the eye-movement-related features. This is a small sample of studies for which to reach any firm conclusions on this topic. However, one possibility may be related to the fact that the participants in the gait-related research were practiced before testing: "…the treadmill training and filming took place after each subject had first walked outdoors, and then inside on the laboratory track." (pg. 67, [15]). It might be worth evaluating the effect of task practice on subsequent test-retest reliability for the eye-movement studies.

In the present case, the reliability of features was assessed in the same sample of subjects used for biometric performance assessment. This is not an ideal situation. In the ideal case, the reliability of features would be established in a distinct sample from that employed in biometric assessment. Also, although the split-sample approach is a highly efficient research tool, testing the biometric performance in completely novel subject samples will eventually be required. Nonetheless, our results illustrate that test-retest reliability is critically important for biometric feature selection, and performance, and that the ICC is an excellent tool to assess it.

## Appendix

## Supplemental Methods

### *A. EM-2 Database*

Short-Term: This database employs 12 fairly simple eye-movement features, 5 of which are simple main sequence measures relevant to saccades [1, 2]. In addition, the feature list includes fixation durations, and mean saccade velocity in the horizontal, and vertical dimensions. Average horizontal, and vertical positions (in degrees of visual angle) are also included. Start times for fixations, and saccades are also extracted. Each of these features has a distribution within each subject, representing the values for each fixation, and each saccade. In the original paper [1], the best results were obtained when the entire distributions were compared. In the present application, we extracted the mean, median, SD, 25th percentile, 75th percentile, skewness, and kurtosis from each of these 12 distributions for each subject. After normalization testing, and redundant feature removal, we had 54 features to test. See Table II (main) for sample sizes, test-retest interval, number of features, number of ICC sets, ICC set sizes, and median ICC.

Long-Term: The data are of exactly the same form as the EM-2-Short-Term database, but only in the long-term subset of subjects (N=68) retested at approximately 11.1 months. See Table II (main) for sample sizes, test-retest interval, number of features, number of ICC sets, ICC set sizes, and median ICC.

### *B. EM- 3 Database*

Short-Term: Dr. Komogortsev, and colleagues have developed an analysis of saccade data, by which estimates can be made of 18 oculomotor plant characteristics (OPC) representing the internal anatomical structure of the eye, including the extraocular muscles, the eye globe, surrounding tissues, and the dynamics of the neuronal control signal [3]. We extracted these 18 features from the same poetry reading database described above. Thus there were 298 subjects, with two sessions each, with 18 features per subject per session. These features were divided into three sets, a HIGH ICC set, a LOW ICC set (above, and below the median), and a third set with all the 18 features (“All”). See Table II (main) for sample sizes, test-retest interval, number of features, number of ICC sets, ICC set sizes, and median ICC.

Long-Term: The data are of exactly the same form as the EM-3-Short-Term database, but only in the long-term subset (N=68) of subjects retested at approximately 11.1 months. See Table II (main) for sample sizes, test-retest interval, number of features, number of ICC sets, ICC set sizes, and median ICC.

### *C. EM-4 Database*

Short-Term: This database is based on horizontal eye movements only and employs 14 fairly simple eye-movement features listed in Supplemental Table I, and described in full in [4]. After normalization testing, and redundant feature removal, we had 10 features to test. See Table II (main) for sample sizes, test-retest interval, number of features, number of ICC sets, ICC set sizes, and median ICC.

Long-Term: The data are of exactly the same form as the EM-3-Short-Term database, but only in the long-term subset (N=68) of subjects retested at approximately 11.1 months. See Table II (main) for sample sizes, test-retest interval, number of features, number of ICC sets, ICC set sizes, and median ICC.

## Supplemental Results

### *A. EM-2 Database*

Short-Term: The analyses for this database are presented in Supplemental Fig. 1, and Table III (main). There was a large range of ICCs (Supplemental Fig. 1, Top). For all sets, for Rank-1-IR, performance generally improved as the number of components extracted increased to 13. For the HIGH ICC set, Rank-1-IR performance was not impressive (Supplemental Fig. 2, Middle, and Table III (main)), but was much higher for the HIGH ICC set than for the LOW set, and the ALL set, even though the ALL set contained twice as many features as the HIGH ICC set, including the HIGH ICC features. For EER, the HIGH ICC set also performed better than the other sets (Supplemental Fig. 1, Bottom, Table III (main)), and achieved its best performance at 10 components, but the level of performance was not very good.

Long-Term: The analyses for this database are presented in Supplemental Fig. 2, and Table III. (main). There is a decrement in the median ICC from the EM-2-Short-Term Database (Median ICC: 0.56 to 0.46), as expected. Despite this, the performance of all the ICC sets is fairly similar to the performance for the EM-2-Short-Term database. Clearly the HIGH ICC dataset performed best for both Rank-1-IR, and EER, and clearly outperformed the “ALL” dataset.

### *B. EM-3 Database*

Short-Term: The analyses for this database are presented in Supplemental Fig. 3, and Table III (main). There are only 18 features for this dataset. There was some variability in the ICC range, but generally, the ICCs are poor for these features (Supplemental Fig. 3, Top, Table II (main)). Neither Rank-1-IR performance nor EER performance were highly dependent on number of components. For the HIGH ICC set, Rank-1-IR performance was very poor (Supplemental Fig. 3, Middle, and Table III (main)), but was higher for the HIGH ICC set than for the ALL set, and especially the LOW ICC set, even though the ALL set contained a twice as many features as the HIGH ICC set, including the HIGH ICC features. For EER, performance changed little with increasing number of components. The HIGH ICC set was narrowly better than the ALL data set but was better than the LOW ICC set (Supplemental Fig. 3, Bottom, Table III (main)). Nonetheless, EER performance with this dataset was quite poor.

Long-Term: The analyses for this database are presented in Supplemental Fig. 4, and Table III (main). The ICCs for this dataset are not very different from the ICCs for the EM-3-Short-Term database (median ICCs: 0.49 vs 0.43). Although the HIGH ICC set outperformed the other sets for both Rank-1-IR, and EER, overall the performance was poor.

| Supplemental Table I. | |
|---|---|
| List of Original Features for EM-4 | |
| **1** | Number of Fixations * |
| 2 | Mean Fixation Duration * |
| 3 | Mean Radial Saccade Amplitude * |
| 4 | Mean Horizontal Saccade Amplitude * |
| 5 | Mean Vertical Saccade Amplitude |
| 6 | Mean Radial Saccade Velocity * |
| 7 | Mean Radial Saccade Peak Velocity * |
| 8 | Slope of the Amplitude/Duration Relationship (saccades) * |
| 9 | Slope of the Main Sequence Relationship |
| 10 | Velocity Waveform Indicator * |
| 11 | Scanpath Length * |
| 12 | Scanpath Complex Hull Area |
| 13 | Regions of Interest |
| 14 | Inflection Count * |
| *Features were normal or normalizable. | |

### *C. EM-4 Database*

Short-Term: The analyses for this database are presented in Supplemental Fig. 5, and Table III (main). There were only 10 total features for this dataset (Table II (main)). The ICCs were generally high (Supplemental Fig. 5, Top). With so few features, PCA was not performed for this database. All 5 features (for HIGH, and LOW), and all 10 features (for ALL) were entered at once for biometric assessment. For Rank-1-IR, and EER, the best performance was for the ALL ICC set (Supplemental Fig. 5, Middle, and Bottom, and Table III (main)). However, even this “best performance” was quite poor.

Long-Term: The analyses for this database are presented in Supplemental Fig. 6, and Table III (main). The median ICC dropped substantially from the EM-4-Short-Term Database (median ICC from 0.79 to 0.46). Rank-1-IR performance was much better for this long-term version of the data, but was still poor. There is not much difference between the short-term, and long-term database for EER.

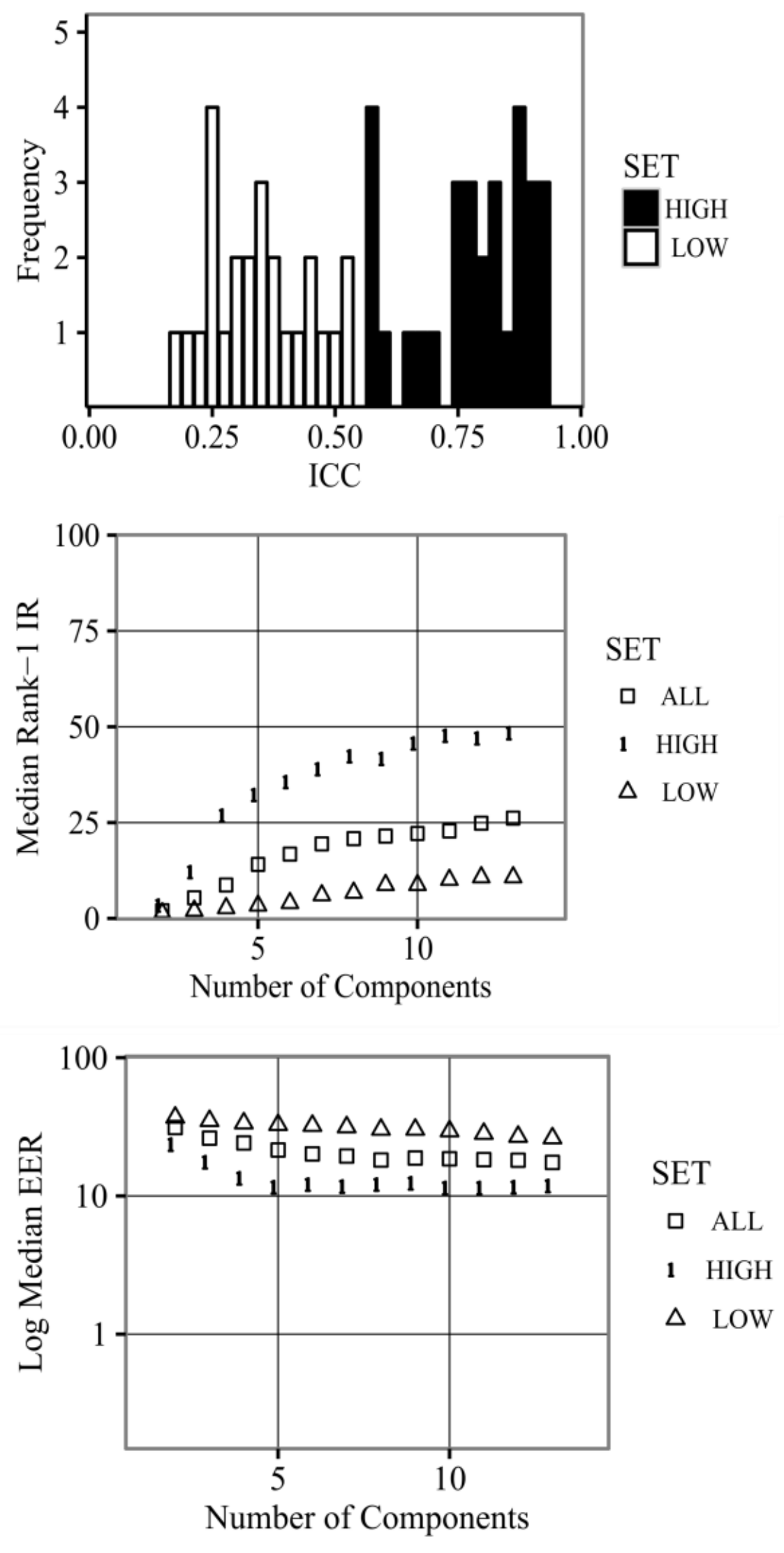

Supplemental Figure 1. Analysis of the EM-2-Short-Term database. See caption for Fig 1. (main).

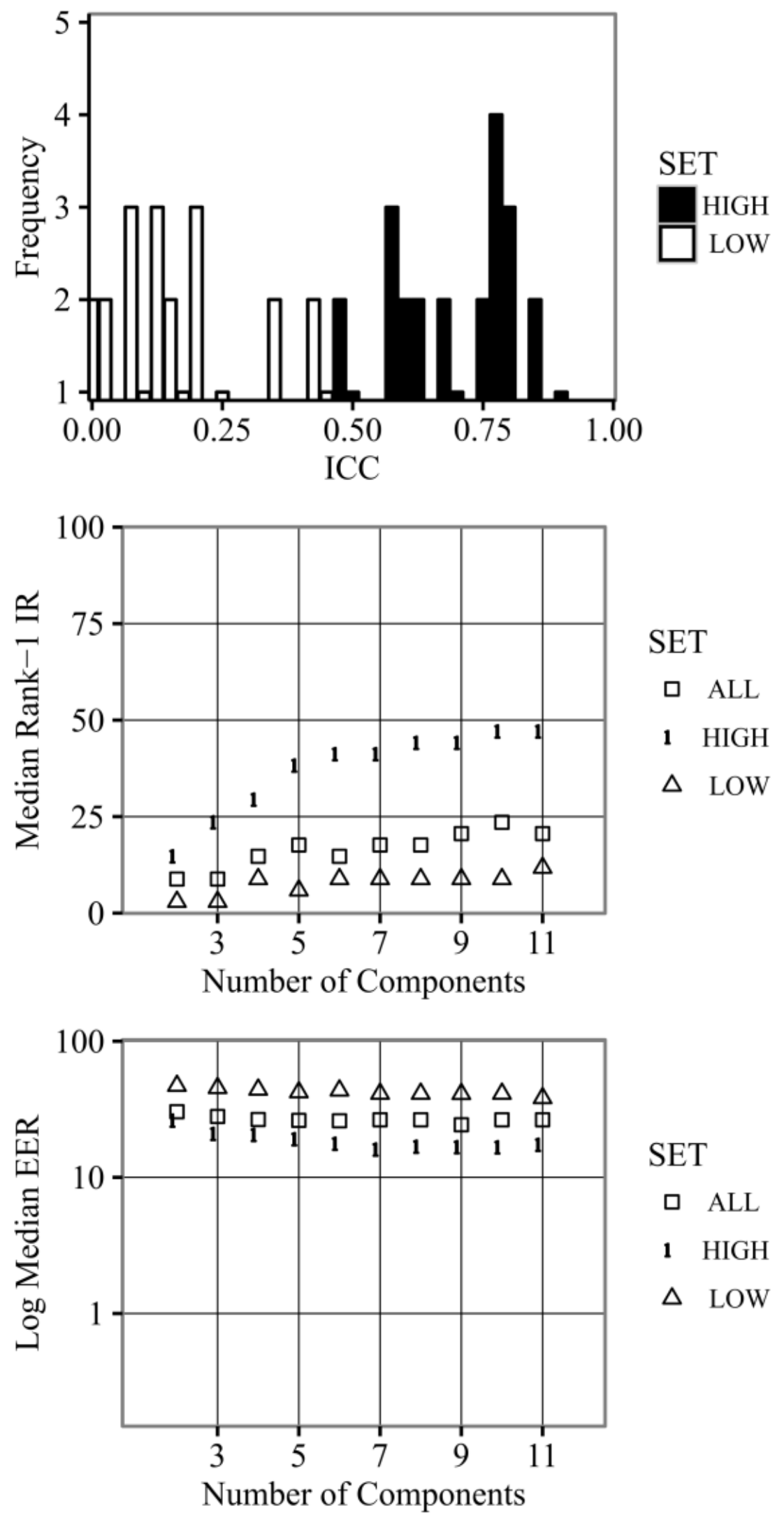

Supplemental Figure 2. Analysis of the EM-2-Long-Term database. See caption for Fig. 1 (main).

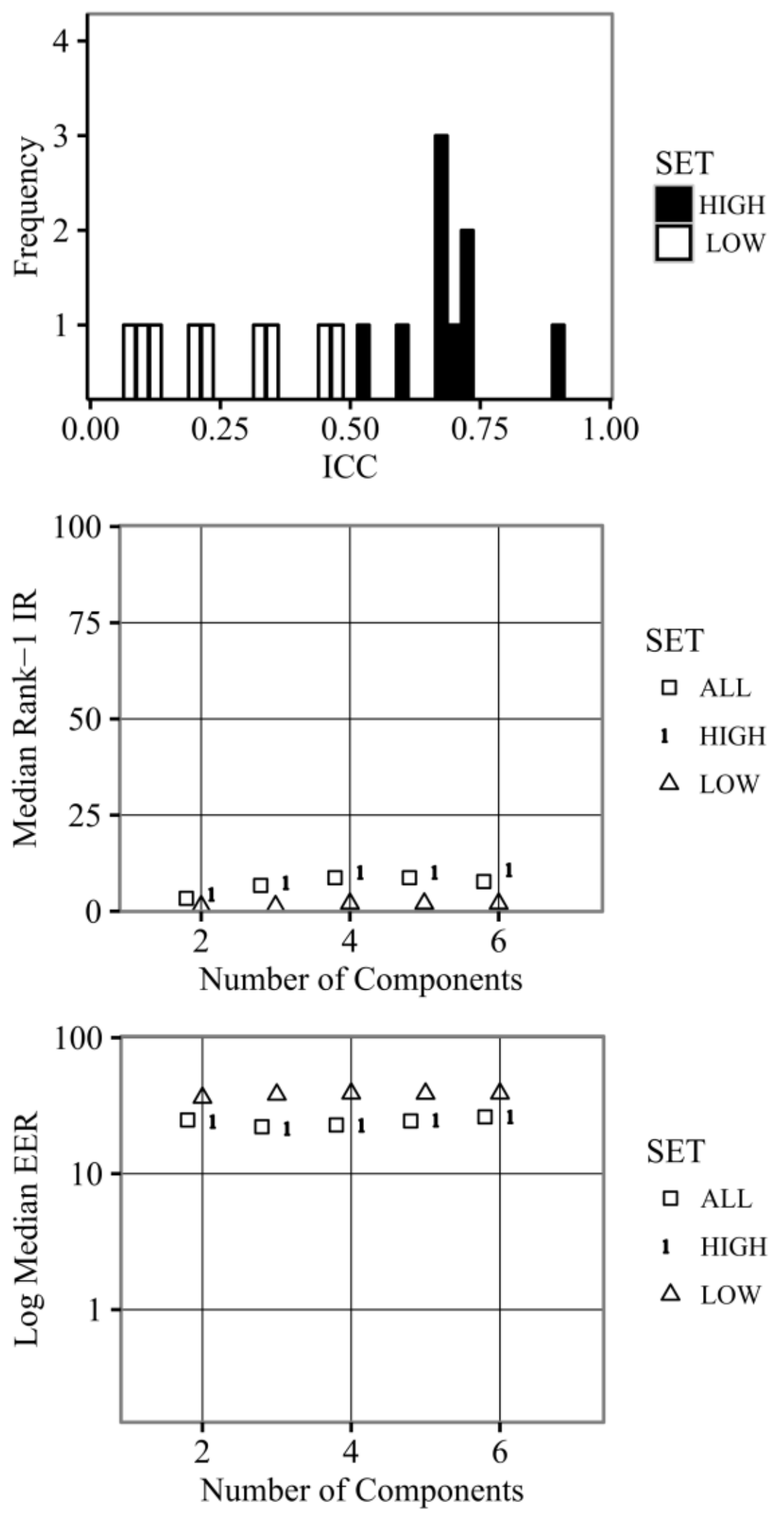

Supplemental Figure 3. Analysis of the EM-3-Short-Term database. See caption for Fig. 1 (main).

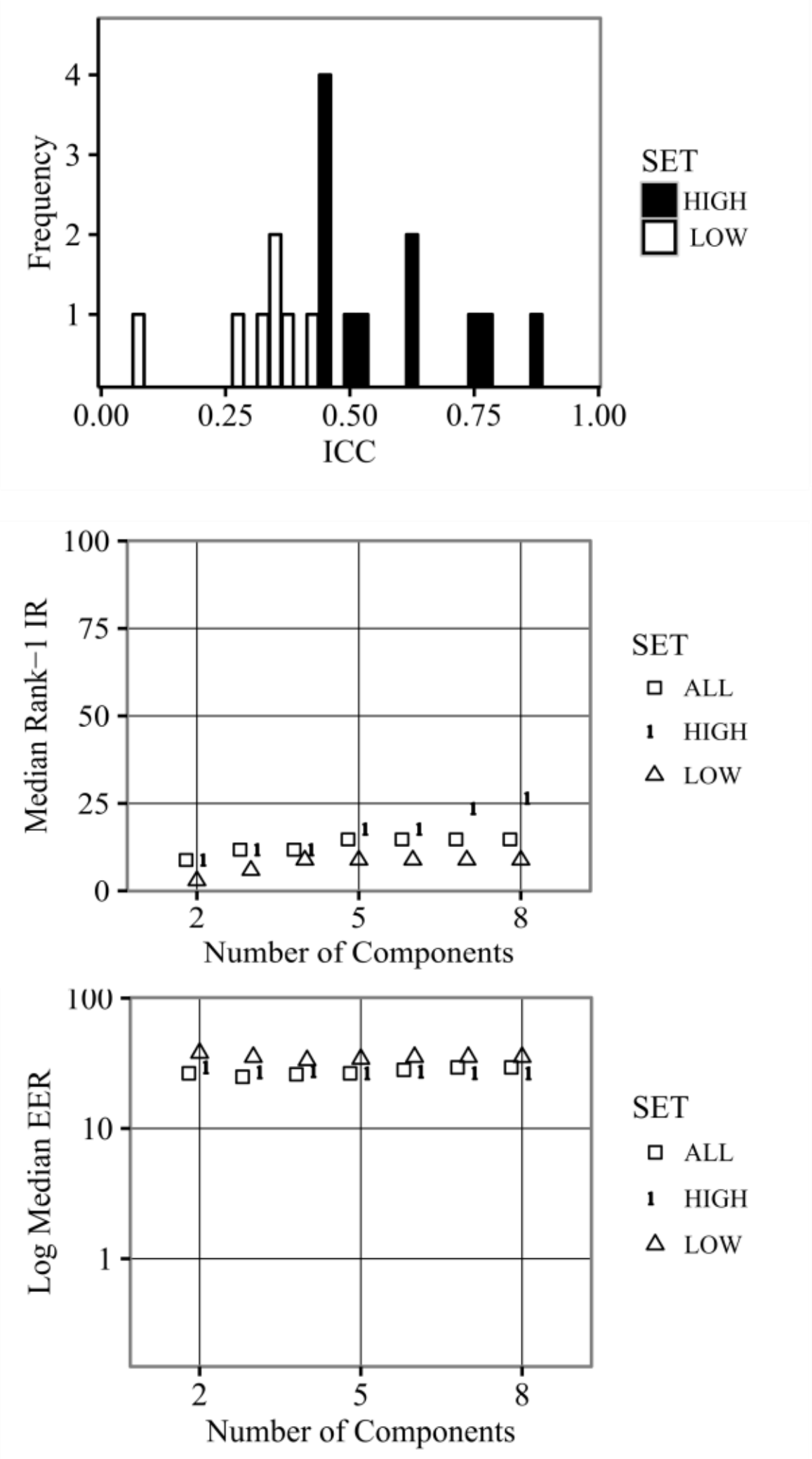

Supplemental Figure 4. Analysis of the EM-3-Long-Term database. See caption for Fig. 1 (main).

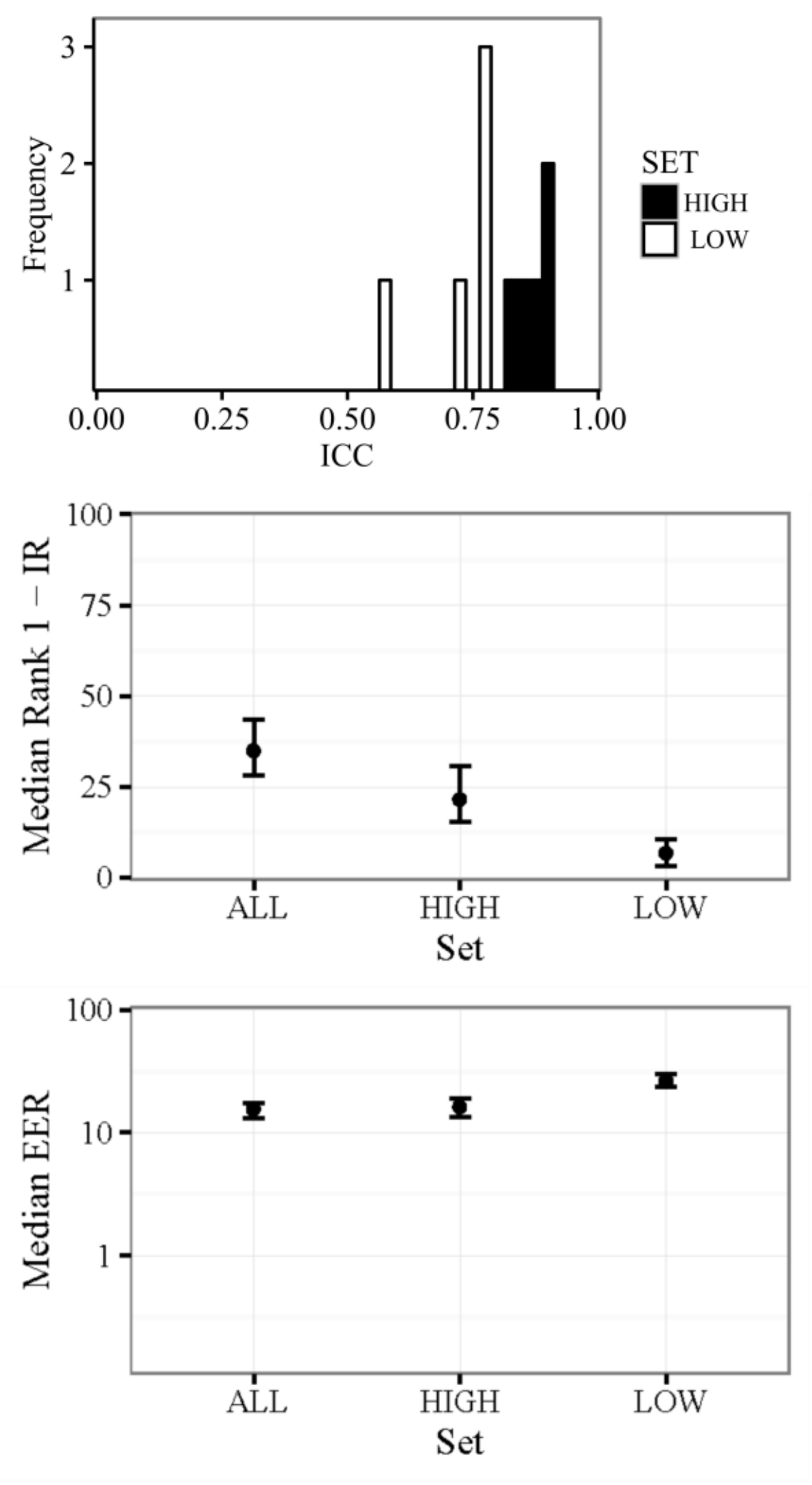

Supplemental Figure 5. Analysis of the EM-4-Short-Term database. PCA was not performed, due to the small number of features. All features were entered directly into the biometric assessment algorithm. Plotted are means (dots) and minimum and maximum value error bars based on 100 training and testing sample sets.

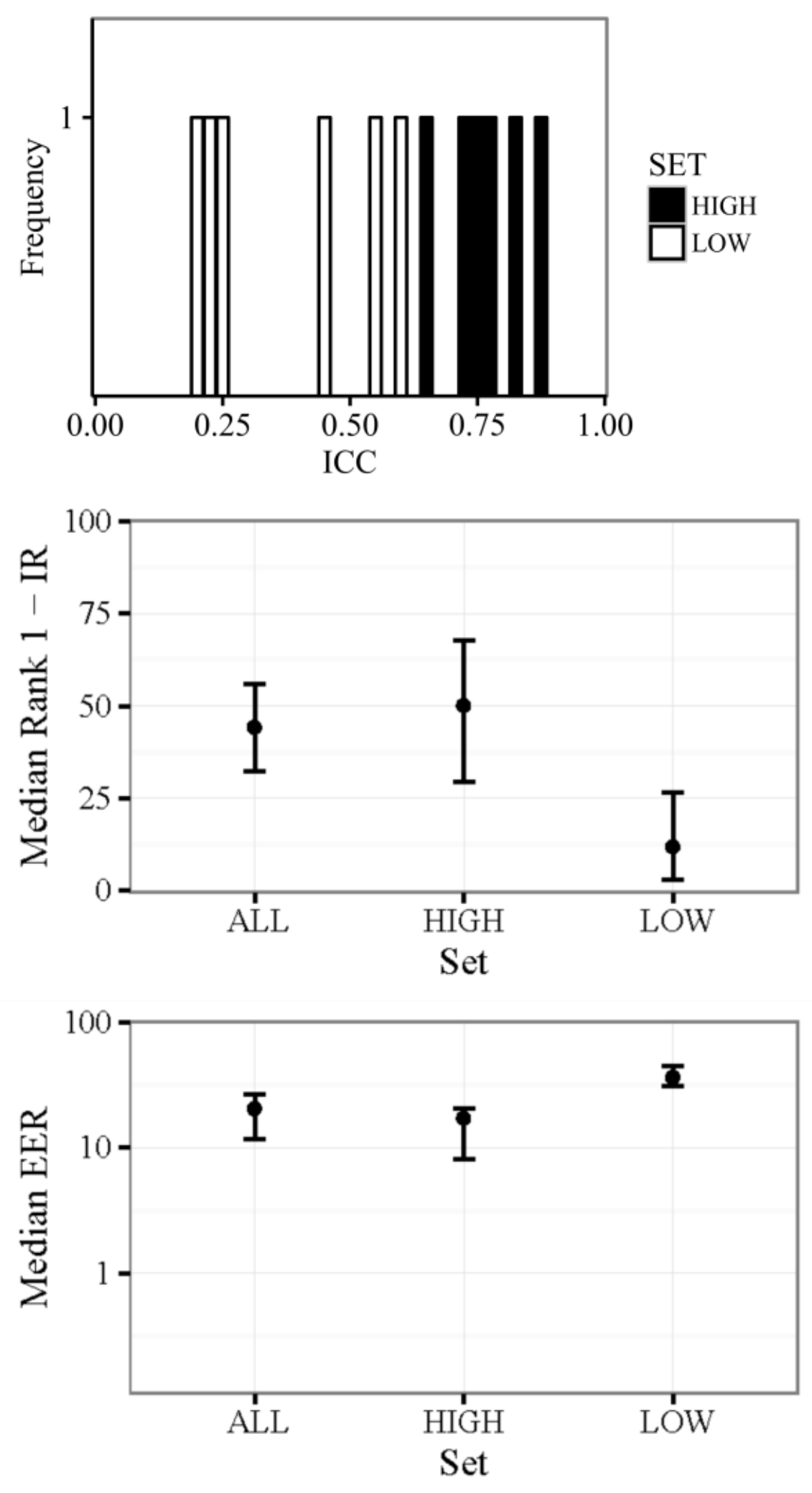

Supplemental Figure 6. Analysis of the EM-4-Long-Term database. See caption for Supplemental Fig. 5.